\documentclass[prb,twocolumn,showpacs,nofootinbib]{revtex4}
\usepackage[dvips]{graphicx}
\usepackage{color,array,dcolumn}
\usepackage{amsmath}
\usepackage{amssymb}

\begin{document}

\title{Quantum Monte Carlo study of circular quantum dots in presence of Rashba interaction}
\author{A. Ambrosetti}
\email{ambrosetti@science.unitn.it}
\affiliation{Dipartimento di Fisica, University of Trento, via Sommarive 14, I--38050, 
Povo, Trento, Italy}
\affiliation{INFN, Gruppo Collegato di Trento, Trento, Italy}
\author{F.Pederiva}
\email{pederiva@science.unitn.it}
\affiliation{Dipartimento di Fisica, University of Trento, via Sommarive 14, I--38050, 
Povo, Trento, Italy}
\affiliation{INFN, Gruppo Collegato di Trento, Trento, Italy}
\author{E. Lipparini}
\email{lipparin@science.unitn.it}
\affiliation{Dipartimento di Fisica, University of Trento, via Sommarive 14, I--38050, 
Povo, Trento, Italy}
\affiliation{INFN, Gruppo Collegato di Trento, Trento, Italy}

\begin{abstract}
\date{\today}
We present the numerical Quantum Monte Carlo results for the ground
state energy of circular quantum dots in which Rashba spin--orbit iteraction
is present. Diffusion Monte Carlo with spin propagation is applied in order to treat
the spin--orbit interaction correctly, following previous work done in the field
of the two--dimensional electron gas. Together with ground state energies, also 
numerical results for density and spin--density profiles are given.
\end{abstract}

\maketitle

\section{Introduction}

Quantum dots have become in the last years very important in several fields, from spintronics
to solar panels and even biological applications. They represent also possible candidates
for qbits realization in view of quantum computation.
The reason for such a success is that they are relatively easy to produce and their discrete energy levels \cite{Kouwenhoven,Steward}
can be engineered depending on specific needs acting over the intensity and shape of the confining
potential.
Though a quantum dot is indeed a quasi zero-dimensional structure, in which electrons are confined along all three dimensions, several possible realizations of quantum dots exist. In this work we consider in particular quantum dots obtained as finite size quasi two--dimensional structures at semicondutor interfaces.
A semiconductor heterostructure provides a
very narrow (001) quantum well from which the quasi two-dimensional character raises. The lateral confinement may then be obtained applying an electrostatic potential or using etching tecniques.
Modern fabrication techniques allow also the control of the number of the confined electrons, which can be very low.

Due to a structure inversion asymmetry of the material an electric field generates inside of the quantum well along the $z$ direction causing
the appearance of the Rashba interaction \cite{Rashba}:
\begin{equation}
V_{Rashba}=\lambda \sum_{i=1}^N [p_i^{y}\sigma_i^{x}-p_i^{x}\sigma_i^{y}] \,, 
\label{vrashba}
\end{equation} 
Here $\mathbf p_i$ is the momentum of the $i-th$ electron, 
and $\sigma_i^x$ and $\sigma_i^y$ are the Pauli matrices acting over 
the spin of particle $i$.
This spin--orbit interaction is particularly interesting due to its tunability \cite{nitta,engels,nittathick}: acting on the gate
voltage it is possible to control the coupling constant $\lambda$. This in principle
gives the possibility of modifying the system spin and the energy levels through a easy to operate
electric potential.  
Spin-orbit effects have a particular interest in confined systems, due to the possible appearance of features like non obvious spin densities and for their effects on energy levels and shell structures.
Several Monte Carlo simulations have already been done in
different conditions \cite{Pederiva,PedeDress,Umrigar}, proving the validity and the accuracy of the method. Other techniques have also
been applied, such as Hartree--Fock, Density Functional--based methods \cite{Reimann} like LDA and LSDA. Also exact
diagonalization \cite{diag1,diag2} was used, but mainly for very small numbers of electrons.
Quantum dots with Rashba interaction were studied in particular in absence of Coulomb potential \cite{Malet} and recently a DFT calculation has been done also taking the Coulomb interaction into account \cite{governale}.
What we propose here is an ab-initio Monte Carlo study, in which the spin-orbit interaction is taken into account by a spin-operator dependent propagator.

\section{Method} 
\label{secmeth}
\subsection{Hamiltonian}
The system studied in this paper is a circular quantum dot with different numbers of electrons trapped
into a static parabolic electric potential of the form $V_{conf}(r)=m\omega_0^2 r^2/2$, where $r=\sqrt{x^2+y^2}$.
The Hamiltonian for such a system with $N$ electrons in presence of the Rashba spin-orbit interaction can be written as
\begin{eqnarray}
H=\sum_{i=1}^N \Big( -\frac{\nabla^2_i}{2m^*} +\frac{m^*}{2}\omega_0^2 r^2_i + \lambda \sum_{i=1}^N (p_i^{y}\sigma_i^{x}-p_i^{x}\sigma_i^{y})  \Big) + \\ \nonumber
 \frac{e^2}{\epsilon}\sum_{i<j}^N \frac{1}{|\mathbf{r_i}-\mathbf{r_j}|}.
\end{eqnarray} 
Here $\epsilon$ and $m^*$ respectively are the effective dielectric constant and electron mass in the semiconductor.
About the confining potential strenght $\omega_0$ and the Rashba interaction coupling constant $\lambda$ different
 values have been used in our calculations in order to obtain a more complete view of the contributions of the two
 interactions to the system ground state.
In the following we will make use of effective atomic units in order to simplify notation, defining $\hbar=e^2/\epsilon=m^*=1$.
The lenght unit therefore is the effective Bohr radius $a_0^*=a_0 \epsilon/m^*$ and the energy unit the effective Hartree $H^*=H \cdot m^*/(m_e\epsilon^2)$. For GaAs quantum dots, we take $\epsilon=12.4$ and $m^*=0.067m_e$, yelding $H^*=11.86meV$ and $a^*_0=97.93 \mathring{A} $.
This Hamiltonian, due to the spin-orbit term, does not commute with the angular momentum operator, but can be proven to commute with the $z$ component of the total momentum $J_z$.

\subsection{Diffusion Monte Carlo}
Diffusion Monte Carlo is a very accurate method used for the study of the ground state properties of many body systems. 
It is based on projection in imaginary time of an initial wavefunction, exploiting the fact that higher energy components of the initial wavefunction will tend to be exponentially suppressed respect to the ground state.
In order to obtain imaginary time evolution of a wavefunction an imaginary time propagator is needed. 
Though an exact form for the propagator is unknown, Trotter's formula is employed in order to write the propagator as a product of exponentials of the different terms of the Hamiltonian. 
This is correct only to second order in the propagation time. The problem is overcome by applying a short time
 propagation repeatedly in order to obtain a long enough projection.
The DMC method samples the ground state wavefunction (multiplied by a trial wavefunction when importance sampling
 is used) with walkers, i.e. points in the phase space moving during propagation. 
The kinetic term of the Hamiltonian produces a gaussian propagator, which is used in DMC as a sampling probability
for walkers displacements. The propagator factor coming from the potential is instead usually seen simply as a
 weight for the walkers.
In case the potential contains terms which only depend on space coordinates this is in principle not a problem.
\newline
 For momentum depending potentials anyway, this approach cannot be applied. Being walkers points in the coordinate
space they don't have a definite momentum, and in almost all interesting cases also the trial wavefunction is
 not a momentum eigenstate.
The Rashba potential, being due to a spin-orbit-like interaction, shows indeed dependence on the particles momenta
 and therefore needs to be treated in a particular way.  
The basic idea (for a detailed description of the method see \cite{io}) is dividing the propagator into
 three factors, containing respectively the kinetic energy, the spin-orbit potential and the remaining parts of
 the Hamiltonian (which then only depend on space coordinates). One can choose to apply the factor containg the
 Rashba potential right after the gaussian free particle propagator. With this procedure the derivative terms  
inside of the spin-orbit interaction are turned into terms with space coordinates and spin operators dependence, having the meaning of a spin rotation. The third factor can finally be applied in the usual way, i.e. as a weight.
The full propagator for the quantum dot in this case can be written as:
\begin{eqnarray}
G(\mathbf{R},\mathbf{R'},\Delta\tau)=e^{-[V_{Coul}(\mathbf{R})+V_{conf}(\mathbf{R})-E_0-\frac{N\lambda^2}{D}]\Delta \tau} \times \\ \nonumber
e^{-i\frac{\lambda}{D}\sum_{i=i}^N(\Delta r^y_i \sigma^x_i -\Delta r^x_i \sigma _i ^y)\Delta \tau} G_0(\mathbf{R},\mathbf{R'},\Delta \tau).
\label{greenfunction}
\end{eqnarray}
With $\mathbf{R}$ and $\mathbf{R'}$ we respectively indicate the set of new and old space coordinates of the N 
electrons, with $V_{Coul}$ the Coulomb potential and with $\Delta r_i^j$ ($j=x,y$) the $x$ and $y$ components of the
difference between the new and the old coordinates of the i-th electron. $G_0$ indicates the free particle propagator
 and $\Delta \tau$ the imaginary time step for the evolution. Notice that the propagator has already been 
renormalized by the factor $\exp(E_0 \Delta \tau)$, where $E_0$ is the ground state energy, in order to let the
ground state weight remain finite.

\subsection{Wavefunction}
Though the quantum dot with parabolic confinement and Rashba interaction can be described in absence of Coulomb potential by a sum
 of single particle Hamiltoninans, its analytical solution has nevertheless not been found yet. The problem one
 encounters when trying to diagonalize such Hamiltonians over a basis of harmonic oscillators, is an infinite 
set of coupled equations. What has been so far proposed is an analytical very accurate approximation and a
 numerical diagonalization over two dimensional harmonic oscillator states \cite{Malet}.
Due to the fact that $[H,J_z]=0$ one expects to get better results for the ground state energy using a trial
 wavefunction which is a $J_z$ eigenstate since such a state would have the required symmetry. Because the
 single particle orbitals deriving from the two methods mentioned above mix different harmonic oscillator states,
 it is difficult building with them a $J_z$ eigenstate. On the other hand one can easily provide $J_z$
 eigenvectors with simple combinations of two dimensional harmonic oscillator orbitals but such
 wavefunctions will be both eigenstates of $S_z$ and $L_z$ and therefore will be eigenstates of operators which
 do not commute with the Hamiltonian. From our Monte Carlo calculations we observed that the wavefunctions
 obtained from simple combinations of harmonic oscillators always give the lowest results for the ground state
 energy, therefore we adopted them obeying to the variational principle.
\newline
Because of the spin rotating factor contained in the imaginary time propagator just shown in the previous section, 
walkers will not only change their space coordinates during propagation, but also their spins. This implies that
 it is not possible separating spin--up and spin--down electrons in two Slater determinants: only one
 determinant must be used containing all electrons. Besides this it was necessary dealing with complex
 numbers because though single particle orbitals could be real, spin components in general are not. For these reasons the trial wavefunction is complex and its phase is not constant. This means that using the fixed 
 node approximation is not possible. Our choice has been the Fixed Phase approximation \cite{io,Bolton,Colletti} which is based on the idea of finding the lowest energy state having the same phase as the trial wavefunction.

\section{Results}
The results contained in this paper concern the ground state of the system for variable numbers of electrons.
 The energy was calculated for two values of the parabolic confining potential and for three values of the Rashba
 interaction strenght in order to give an extensive and systematic study of the system. 
Studying the system under different conditions is meant to partially account for the tunability of the two potentials.

\begin{table}
\begin{tabular}{ccccccc}
\multicolumn{7}{c}{} \\
\hline 
\hline
  &   &     &          &$\omega=0.28$&           &          \\
\hline 
\hline
$N_{el}$ & L & S & $\lambda=0$ & $\lambda=0.1$ & $\lambda=0.35$ & $\lambda=0.7$ \\
\hline
2 & 0 & 0   & 0.51081(3) & 0.5118(3) & 0.5120(3) & 0.5153(3)\\
3 & 1 & 1/2 & 0.7446(2)  & 0.746(1)  & 0.748(1)  & 0.753(1) \\
4 & 0 & 0   & 0.94520(5) & 0.944(1)  & 0.946(1)  & 0.951(1) \\
4 & 0 & 1   & 0.92863(5) & 0.934(1)  & 0.934(1)  & 0.939(1) \\
5 & 1 & 1/2 & 1.1067(2)  & 1.117(1)  & 1.118(1)  & 1.127(1) \\
6 & 0 & 0   & 1.2666(4)  & 1.282(1)  & 1.285(1)  & 1.289(1) \\
7 & 2 & 1/2 & 1.4337(4)  & 1.457(1)  & 1.463(1)  & 1.493(3) \\
\hline
\hline

  &   &     &          &$\omega=1.78$&           &          \\
\hline
\hline
$N_{el}$ & L & S & $\lambda=0$ & $\lambda=0.1$ & $\lambda=0.35$ & $\lambda=0.7$ \\
\hline
2 & 0 & 0   & 2.4820(6)  & 2.4805(6) & 2.4803(5) & 2.4815(5) \\
3 & 1 & 1/2 & 3.469(1)   & 3.4663(4) & 3.4669(5) & 3.485(5)  \\
4 & 0 & 0   & 4.200(2)   & 4.1968(4) & 4.1969(4) & 4.1975(4) \\
4 & 0 & 1   & 4.150(1)   & 4.151(1)  & 4.148(1)  & 4.150(1)  \\
5 & 1 & 1/2 & 4.784(1)   & 4.7832(5) & 4.7839(5) & 4.786(1)  \\
6 & 0 & 0   & 5.332(2)   & 5.326(1)  & 5.326(1)  & 5.329(1)  \\
7 & 2 & 1/2 & 6.003(1)   & 6.004(1)  & 6.003(1)  & 6.012(2)  \\
\hline
\hline
\end{tabular}
\label{tabenergy}
\caption{Diffusion Monte Carlo results for the ground state energy per particle for different numbers of electrons and values of the Rashba coupling constant $\lambda$. Results are reported for two different confinement potential streghts. Energies are expressed in effective Hartrees ($H^*$). }
\end{table}

\begin{figure}[ht]
\vspace{0.5cm}
\centering
\includegraphics[scale=0.28]{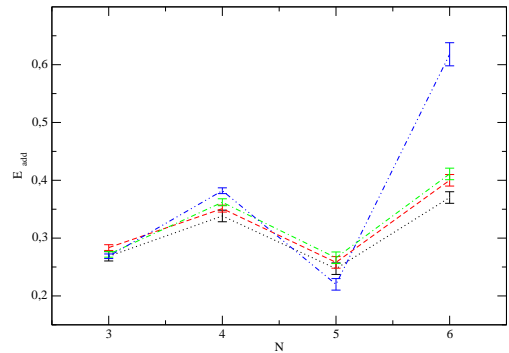}
\caption{(color online) Addition energy for $\omega=0.28$. $\lambda=0$ are represented by the line with dots (black) ,$\lambda=0.1$ dashes (red) ,$\lambda=0.35$ dot--dash (green), $\lambda=0.7$ double dot--dash (blue). Data are given in effective atomic units.}
\label{add1}
\end{figure}

\begin{figure}[ht]
\vspace{0.5cm}
\centering
\includegraphics[scale=0.28]{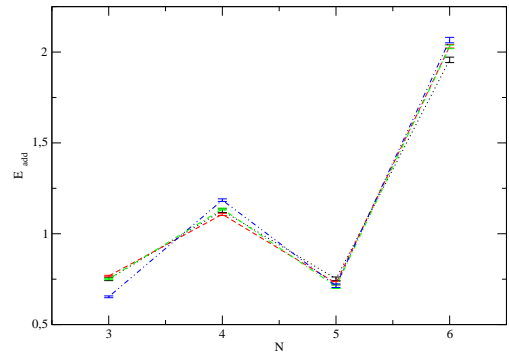}
\caption{(color online) Addition energies for dots with confinement strenght $\omega=1.78$  $\lambda=0$ dots (black),$\lambda=0.1$ dashes (red),$\lambda=0.35$ dot--dash (green),$\lambda=0.7$ double dot--dash (blue). Data are given in effective atomic units.}
\label{add2}
\end{figure}

The results reported in table \ref{tabenergy} show the different behavior of the
system depending on the value of $\omega$. For smaller confinement it is clear how
the spin orbit interaction always increases the ground state energy. Moreover the energy always 
shows the same increasing trend with increasing $\lambda$.
Something different happens for higher values of the parabolic confinement. Though, due to 
errorbars it is not possible giving a detailed picture, ground state energies can be lower
in presence of the spin orbit interaction, in particular for low values of the Rahsba coupling
constant. They tend in any case to increase for high Rashba potential strenghts.
\newline
In Fig. \ref{add1},\ref{add2} the addition energies for the system are
reported respectively for $\omega=0.28$ and $\omega=1.78$.
The addition energy is defined as
\begin{equation}
E_{add}=\mu (N+1)- \mu (N)
\end{equation}
where $\mu (N)$ is the chemical potential of a $N$ electrons quantum dot.
In both cases it is possible noticing that the behavior of $E_{add}$ as
a function of the electrons number becomes consistently steeper with increasing $\lambda$.
At $\omega=1.78$ comparison is possible with data from LSDA calculations \cite{governale}. Such comparison
gives resonable agreement for $\lambda=0$ but show different dependence on the Rashba coupling constant. Deviations become bigger increasing $\lambda$ since LSDA predicts a change in the shell structure.
\newline

\begin{figure}[ht]
\vspace{0.5cm}
\centering
\includegraphics[scale=0.28]{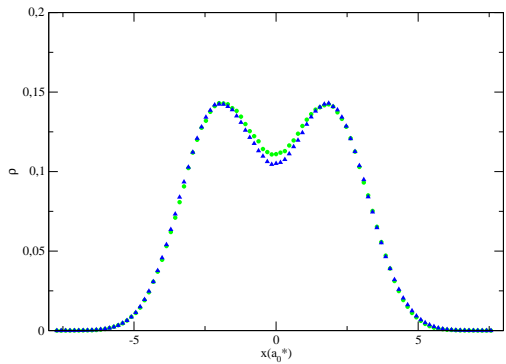}
\caption{(color online) Density of a $6$ electons dot with confinement strenght $\omega=0.28$ in presence of and without Rashba interaction. Triangles (blue) refer to results with $\lambda =0$ while circles (green) represent results at $\lambda=0.35$. Data are given in effective atomic units.}
\label{denscomp}
\end{figure}

\begin{figure}[ht]
\vspace{0.5cm}
\centering
\includegraphics[scale=0.6,angle=0]{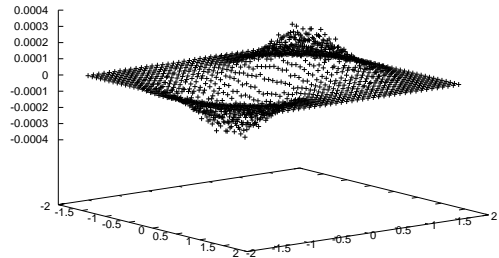}
\caption{Rotation--averaged Y Spin Density of a $2$ electons dot with confinement strenght $\omega=1.78$ in presence of Rashba interaction $\lambda=0.1$. Data are reported in effective atomic units ($a_0^*$ for lenghts) and $(a_0^*)^{-2}$ for the density.}
\label{xdens}
\end{figure}

Quantum dots are known to show interesting density profiles, changing shape with the electrons number \cite{Reimann}. In particular, a $6$ electrons dot shows a little $hole$ in the density profile in its center. Such distributions are due to the interplay between the Coulombic repulsion and the kinetic energy within the confining potential.
From Fig. \ref{denscomp} it is possible noticing how the Rashba interaction is effective in modifying the density of the dot, rendering the central $hole$ shallower.
As one might expect the spin orbit interaction does not allow the occupation of $S_z$ eigenstates since $[H,S_z]$ is non zero once the interaction is switched on. This implies a different effect on average of the Coulomb potential and at the same time a modification of the kinetic energy.
\newline
Within the QMC approach it is possible sampling, apart from the density, also the spin density of the system, also studied within the DFT approach \cite{Reimannlett}. This correspond to the following expectation value:
\begin{equation}
\rho_j(\mathbf{r})=\frac{<\phi(\mathbf{r'})|\sum_i^N\delta(\mathbf{r}-\mathbf{r'})\sigma_i^j|\phi(\mathbf{r'})>}{<\phi(\mathbf{r'})|\phi(\mathbf{r'})>},
\end{equation}
with $j=x,y,z$.
From Monte Carlo simulations such quantities in the two electrons case always gave fluctuating results, though always showing the same kind of structure in all cases ($x,y,z$). Such structures appeared with different orientations and peaks heights. This suggests that such spin densities are more related to correlation effects and should be connected to some two body operator.
The results reported in Fig. \ref{xdens} show the average of 150 spin density results rotated in order to obtain alignment of the common structure along the $x$ axis. 
From such figure it is possible noticing how the two different spin components tend to arrange in opposite sides of the dot.
The structures of Fig. \ref{xdens} are common to all three spin density components ($x,y,z$) which give the same average results within errorbars (not reported in the graph, but of the order of 20 percent).

\begin{figure}[ht]
\vspace{0.7cm}
\centering
\includegraphics[scale=0.6,angle=0]{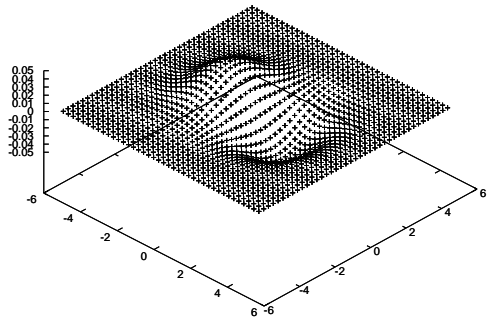}
\caption{The plot shows the X Spin Density of a $3$ electons dot with confinement strenght $\omega=0.28$ in presence of Rashba interaction $\lambda=0.1$. Data are reported in effective atomic units ($a_0^*$ for lenghts) and $(a_0^*)^{-2}$ for the density.}
\label{3densx}
\end{figure}

\begin{figure}[ht]
\vspace{0.7cm}
\centering
\includegraphics[scale=0.6,angle=0]{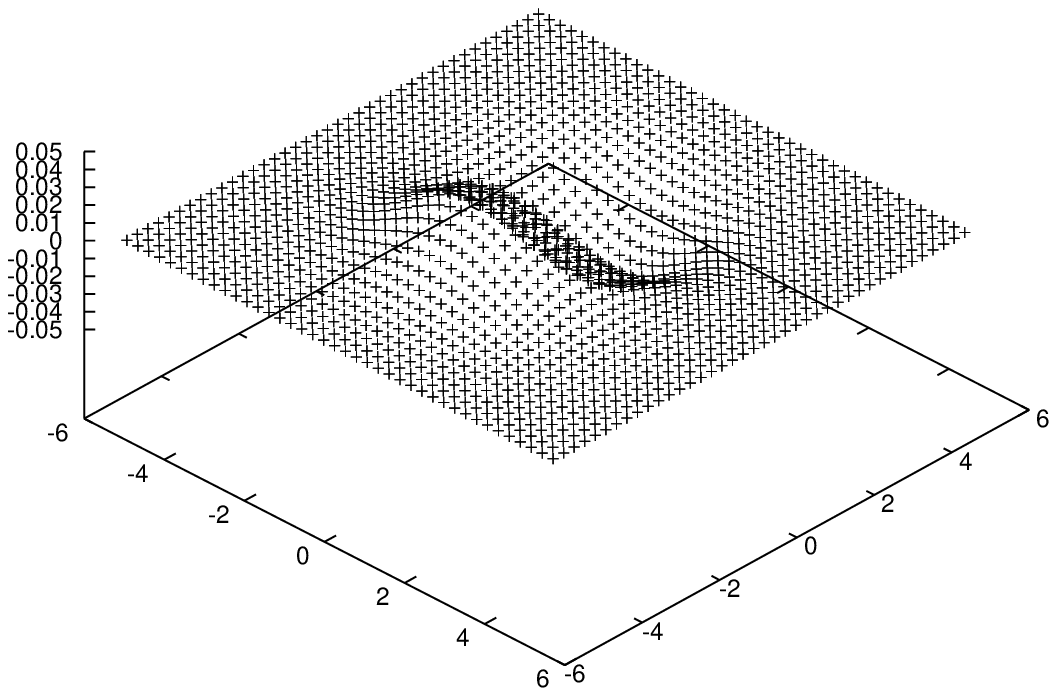}
\caption{Y Spin Density of a $3$ electons dot with confinement strenght $\omega=0.28$ in presence of Rashba interaction $\lambda=0.1$. Data are reported in effective atomic units ($a_0^*$ for lenghts) and $(a_0^*)^{-2}$ for the density.}
\label{3densy}
\end{figure}

\begin{figure}[ht]
\vspace{0.7cm}
\centering
\includegraphics[scale=0.6,angle=0]{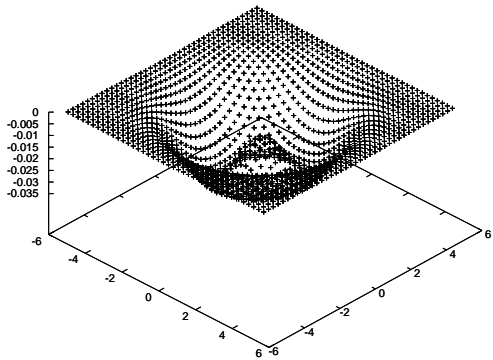}
\vspace{0.5cm}
\caption{Z Spin Density of a $3$ electons dot with confinement strenght $\omega=0.28$ in presence of Rashba interaction $\lambda=0.1$. Data are reported in effective atomic units ($a_0^*$ for lenghts) and $(a_0^*)^{-2}$ for the density.}
\label{3densz}
\end{figure}

Calculations for the spin-density were done also for a 3 electrons quantum dot with $\omega=0.28$ and $\lambda=0.1$. While in the 2 electrons system a non rotated average would give results consistent with zero spin-density within the errorbars, the 3 electron system shows interesting spin patterns and non zero magnetization. This is due to the fact that, while a two electron dot has a closed shell configuration, the addition of a third electron causes the partial occupation of the second shell, corresponding to non zero angular momentum and spin.
From fig. \ref{3densx},\ref{3densy},\ref{3densz} it is clear how the $z$ spin density has a structure which strongly differs from the $x$ and $y$ spin densities. In particular, within errorbars, which are of the order of $10$ percent, the $z$ spin density has circular symmetry, while the $x$ has a structure very similar to the $y$ spin density, though having orthogonal orientation.
 It is interesting to mention how such structures show a general resemblance to the analytical results one would obtain in absence of Coulomb interaction using approximate solutionsi \cite{valin1,valin2}. Since the trial wave function employed in QMC calculations was a combination of gaussian orbitals and carried no information about the effects of the spin orbit interaction on the system, we stress that such peculiar structures are only due to the spin orbit propation given by the Green's function \eqref{greenfunction}.

\section{Conclusions}
This study represents an application of the DMC method with spin--orbit propagator \cite{io} in the direction of confined electrons system. It opens therefore the way to further applications of Diffusion Monte Carlo to nanosystems and molecules in presence of spin--orbit interactions which should not necessarily coincide with the Rashba interaction.
The different structure of the energy levels depending on the confining potential suggests the possibility of having more freedom in the engineering of energy levels. 
Though, due to errorbars a full characterization of the energy levels structure of the system is not yet possible, the results presented show nevetheless an interesting and non trivial dependence on the two parameters $\lambda$ and $\omega$.
The method allowed furthermore the study of spin orbit effects on the electron density and spin density.

\section{acknowledgements}
Calculations were performed on the Wiglaf cluster of the Physics department of the University of Trento and on CINECA computers under grants of the Univeristy of Trento.
We want to thank Dr. Francesc Malet for useful discussion.

\end{document}